\documentclass[aps,prd,twocolumn,floats,floatfix,nofootinbib]{revtex4-1}
\usepackage{graphicx}
\usepackage{amssymb}
\usepackage{epstopdf}
\usepackage{hyperref}
\usepackage{multirow}
\usepackage{amsmath}

\usepackage{amsmath}
\usepackage{bbold}
\usepackage{color}

\begin{document}

\title{Dark Matter freezeout in modified cosmological scenarios}

\author{Alexandre Poulin}
\email{apoulin@physics.carleton.ca}

\affiliation{Ottawa-Carleton Institute for Physics, Carleton University, 1125 Colonel By Drive, 
Ottawa, Ontario K1S 5B6, Canada}

\date{\today}                                  

\begin{abstract}
We study the effects of modifying the expansions history of the Universe on Dark Matter freezeout. We derived a modified Boltzmann equation for freeze-out for an arbitrary energy density in the early Universe and provide an analytic approach using some approximations. We then look at the required thermally averaged cross sections needed to obtain the correct relic density for the specific case where the energy density consists of radiation plus one extra component which cools faster. We compare our analytic approximation to a numerical solutions. We find that it gives reasonable results for most of the parameter space explored, being at most a factor of order one away from the measured value. We find that if the new contribution to the energy density is comparable to the radiation density, then a much smaller cross section for Dark Matter annihilation is required. This would lead to weak scale Dark Matter being much more difficult to detect and opens up the possibility that much heavier Dark Matter could undergo freezeout without violating perturbative unitarity.
\end{abstract}

\maketitle

\section{Introduction}

Dark Matter is still one of the biggest unsolved puzzles in modern physics. It continues to escape detection in both direct and indirect detection experiments while collider experiments have yet to be able to identify a statistically significant Dark Matter signal. If one wishes to construct a particle physics model which can explain Dark Matter, then they must ensure that its interactions are quite weak to avoid all these constraints. For recent reviews on the topic of Dark Matter, see \cite{Bertone:2004pz,Feng:2010gw,Gelmini:2015zpa,Roszkowski:2017nbc,Arcadi:2017kky,Battaglieri:2017aum}

One measurement we do have is the Dark Matter relic density, $\Omega_{CDM}h^2=0.120\pm 0.001$ \cite{Aghanim:2018eyx}. This measurement has been a key part of model building as a robust model must be able to explain this observation with some production mechanism. A popular production mechanism that has been studied is freeze-out. In this scenario, the Dark Matter starts in thermal equilibrium with the Standard Model at an early time but eventually decouples once the rate of the reactions maintaining chemical equilibrium become comparable to the expansion rate of the universe \cite{Bertone:2004pz,Feng:2010gw,Gelmini:2015zpa,Roszkowski:2017nbc,Arcadi:2017kky,Battaglieri:2017aum,Kolb:1990vq}. This is the usual mechanism used for Weakly Interacting Dark Matter (WIMPs) models \cite{Bertone:2004pz,Feng:2010gw,Gelmini:2015zpa,Roszkowski:2017nbc,Arcadi:2017kky,Battaglieri:2017aum,Kolb:1990vq}.

Ensuring that freezeout produces the right amount of Dark Matter can fix some of the couplings in a model. This can lead to a model of Dark Matter being ruled out if the required couplings are too big to simultaneously avoid experimental bounds while producing enough Dark Matter. This has led to models which add more free parameters such as additional interactions between the Dark sector and the Standard Model, including multiple stable Dark Matter species, or considering a different production mechanisms all together.

However, WIMPs are very popular because of the so-called WIMP miracle where one could achieve the correct relic density through freeze-out with weak scale masses and couplings. Many extensions to the Standard Model naturally or can easily accommodate a WIMP like particle. Because of this, it is important to understand how freezeout works and how modifications to the standard cosmological picture could modify the freeze-out results. 

In the standard model of cosmology, freezeout occurs during the radiation dominated era. However, because we do not have any observations before Big Bang Nucleosynthesis (BBN), we cannot say for sure what the expansion history of the universe was before BBN. Namely, we cannot exclude the possibility of an additional contribution to the energy density that cools at a faster rate than radiation. This can come about from alternative cosmological models like alternative Dark Energy models \cite{Linder:2002et,vonMarttens:2018iav}, anisotropic expansion \cite{Barrow:1982ei,Kamionkowski:1990ni}, brane world cosmology \cite{Stoica:2000ws,Okada:2004nc,AbouElDahab:2006glf,Guo:2009nt,Iminniyaz:2015wva}, some inflaton models \cite{Arbey:2008kv,Arbey:2011gu,Bastero-Gil:2015lga}, some quintessence models \cite{Salati:2002md,Rosati:2003cu,Profumo:2003hq,Pallis:2005hm,Iminniyaz:2013cla}, or scalar-tensor gravity \cite{Catena:2004ba,Catena:2006bd}. There has also been work on early matter dominated eras which decay away \cite{Bernal:2018ins,Bernal:2018kcw,Hamdan:2017psw,Hardy:2018bph}. All these models have a different expansion rate from the standard $\Lambda_{\rm CMD}$ model and some previous work has been done to understand the physics in these scenarios \cite{DEramo:2017gpl,DEramo:2017ecx,Maldonado:2019qmp,Iminniyaz:2018das}.

In this work, we develop a general freeze-out equation that can be used for any modified expansion rate and includes the effects of changes in entropy density. In section \ref{sec:derivation}, we go through the derivation of the changes to the Boltzmann equations. In section \ref{sec:Analytic}, we take an analytic approach to obtain a simple way to approximate the relic density. In section \ref{sec:numAna}, we look at numerical results from solving the modified Boltzmann equations for the case of a simple change to the energy density and compare these results to our approximations. Finally, in section \ref{sec:conclusions}, we summarize our conclusions.

\section{Derivation of the Modifications to the Boltzmann equation}
\label{sec:derivation}

In this section, we  derive changes to the Boltzmann equation based on modification to the Hubble Parameter. Many of the definitions and approximations are the same as those of Kolb and Turner \cite{Kolb:1990vq}, namely CP conservation of the matrix elements, no Fermi degeneracy, no Bose-Einstein condensates, and that the temperature of the Dark Matter will be the same the photon temperature until after freeze-out.

\subsection{Single effective component}

In this section, we will derive the Boltzmann equation for a general energy density $\rho(T)$ where $T$ is the photon temperature. We will also use the equation of state $p(T)=w(T)\rho(T)$ which relates the energy density to the pressure $p(T)$. It will be important to not consider $w(T)$ as simply a constant for this treatment. This energy density can be any well-behaved function of temperature. We start with the Friedman equation:
\begin{align}
H^2=\frac{8\pi G}{3}\rho(T), \label{eq:FriedmanSingle}
\end{align}
where $G$ is Newton's constant. To take the expansion of the universe into account, it is typical to use the entropy density as a fiducial quantity. It is defined as:
\begin{align}
s=\frac{\rho+p}{T}=\frac{\rho(1+w)}{T}. \label{eq:s}
\end{align}
We will assume that the there is no entropy injection during freezeout so that the entropy per comoving volume $S$ is conserved. This results in in $s\propto a^{-3}$ where $a$ is the scale factor, which gives $\dot{s}=-3Hs$. 

We now introduce the dimensionless quantity $x=m/T$ where $m$ is some appropriate mass scale, typically the mass of the Dark Matter species. It will be important to find an expression for $\dot{x}$. This can be done by looking at $\dot{s}$:
\begin{align}
\dot{s}&=\frac{\dot{x}}{m}\left(\rho(1+w)+x(1+w)\frac{d\rho}{dx}+x\rho\frac{dw}{dx}\right),\label{eq:sdot}\\
\dot{x}&=-3xH\left(1+\frac{d\log\rho}{d\log x}+\frac{d\log(1+w)}{d\log x}\right)^{-1}. \label{eq:xdot}
\end{align}

The usual Boltzmann equation for a single species is written as \cite{Kolb:1990vq}:
\begin{align}
\dot{n}+3Hn=&-\langle\sigma v \rangle \left(n^2-n^2_{\rm eq}\right)\label{eq:boltz},
\end{align}
where $n$ is the number density, $\langle\sigma v \rangle$ is the thermally-averaged DM annihilation cross-section times velocity, and $n_{\rm eq}$ is the equilibrium number density given by \footnote{For the case of asymmetric Dark Matter, one must include non-zero chemical potentials in the exponential. See \cite{Kolb:1990vq,Iminniyaz:2018das} for details.} :
\begin{align}
n_{\rm eq}=\frac{gm^2T}{2\pi^2}K_2\left(\frac{m}{T}\right)\approx g\left(\frac{mT}{2\pi}\right)^{3/2}e^{-m/T}, \label{eq:neq}
\end{align}
where $g$ is the number of internal degrees of freedom, $m$ is the particles mass, and $K_2$ is the modified Bessel function of the second kind. The approximation is valid for large $x$ or small $T$. \footnote{It is typical to choose the mass scale in the definition of $x$ to be the same as the mass of the particle as it simplifies some of these equations. However, in the case of multi-species Dark Matter models where not all the species are the same mass, one must distinguish the mass scale in the definition of $x$ from the mass in the definition of $n_{\rm eq}$ \cite{Poulin:2018kap}.}

We now make the substitutions 
\begin{align}
Y=&\frac{n}{s},\\
Y_{\rm eq}=&\frac{n_{\rm eq}}{s}.
\end{align}
The derivative of $Y$ with respect to time is:
\begin{align}
\dot{Y}=&\frac{\dot{n}}{s}-\frac{n}{s^2}\dot{s},\\
s\dot{x}\frac{dY}{dx}=&\dot{n}+3nH, \label{eq:ydot}
\end{align}
where we used $\dot{s}=-3Hs$. Combining Eq. \ref{eq:xdot}, \ref{eq:boltz}, and \ref{eq:ydot}, we obtain:
\begin{widetext}
\begin{align}
\frac{dY}{dx}=&-\frac{s}{\dot{x}}\langle\sigma v \rangle \left(Y^2-Y^2_{\rm eq}\right),\label{eq:generalSimple}\\
\frac{dY}{dx}=&\left(\frac{\rho}{24\pi m^2 G}\right)^{1/2}(1+w)\left(1+\frac{d\log\rho}{d\log x}+\frac{d\log(1+w)}{d\log x}\right)\langle\sigma v \rangle \left(Y^2-Y^2_{\rm eq}\right). \label{eq:general}
\end{align}
\end{widetext}

Eq. \ref{eq:general} can be used for any function of $\rho$ and $w$. It is currently written for the case of a single Dark Matter species, but extending it to the case of multiple Dark Matter species is straight forward. See \cite{DHT2018,Poulin:2018kap} for more details on the multi-species case.

\subsection{Multiple energy densities}

Equation \ref{eq:general} is valid for any initial $\rho(T)$. However, in the case where one simply adds an extra contribution to the energy density, we can write the energy density as multiple contributing energy densities which is not only simpler than treating $w$ as a function of temperature, but also makes the equation clearer. In this section, we look at the case where we can write $\rho=\sum_i \rho_i$ with $p_i=w_i\rho_i$ with $w_i$ all constant. Following the same steps used to obtain Eq. \ref{eq:xdot}, we find
\begin{align}
\frac{s}{\dot{x}}&=-\frac{1}{3mH}\sum_i\rho_i(1+w_i)\left(1+\frac{d\log\rho_i}{d\log x}\right).\label{eq:sbyxdotComp}
\end{align}
Combining Eq \ref{eq:boltz}, \ref{eq:ydot}, and \ref{eq:sbyxdotComp}, we obtain:
\begin{widetext}
\begin{align}
\frac{dY}{dx}=&\left(\frac{1}{24\pi m^2 G(\sum_i\rho_i)}\right)^{1/2}\left(\sum_i\rho_i(1+w_i)\left(1+\frac{d\log\rho_i}{d\log x}\right)\right)\langle\sigma v \rangle \left(Y^2-Y^2_{\rm eq}\right). \label{eq:generalComp}
\end{align}
\end{widetext}

The radiation energy density in the early universe is given by:
\begin{align}
\rho_r=g_*(x)\frac{\pi^2}{30}\left(\frac{m}{x}\right)^4, \label{eq:rhor}
\end{align}
where $g_*(x)$ is the number of relativistic degrees of freedom. If we set $\rho=\rho_r$ in Eq \ref{eq:generalComp}, we obtain the usual result
\begin{align}
\frac{dY}{dx}=&-\left(\frac{\pi g_*(x)}{45 G }\right)^{1/2}\left(1-\frac{1}{3}\frac{d\log g_*}{d\log x}\right)\frac{m\langle\sigma v \rangle}{x^2} \left(Y^2-Y^2_{\rm eq}\right).\label{eq:radOnly}
\end{align}
The $\frac{d\log g_*}{d\log x}$ term is usually ignored because it is small, but we include it in our analysis for completeness.

As a final example, we follow the example from \cite{Iminniyaz:2018das}. Let 
\begin{align}
\rho=\rho_r+\rho_D \left(\frac{T}{T_0}\right)^{n_D}=\rho_r\left(1+\frac{g_*(x_0)}{g_*(x)}\eta\left(\frac{x_0}{x}\right)^{n_D-4}\right),
\end{align}
where $\rho_D$, $n_D>4$, and $\eta=\frac{\rho_D(T_0)}{\rho_r(T_0)}$ are all constants, $T_0$ is some constant reference temperature and $x_0=m/T_0$. In our analysis, we will assume that $p_D=w_D\rho_D$ with $3(1+w_D)=n_D$. Putting this into a similar form as Eq. \ref{eq:radOnly} which only included radiation, we obtain:
\begin{widetext}
\begin{align}
\frac{dY}{dx}=&-\left(\frac{\pi g_*(x)}{45 G \left(1+\frac{g_*(x_0)}{g_*(x)}\eta\left(\frac{x_0}{x}\right)^{n_D-4}\right)}\right)^{1/2}\left[1-\frac{1}{3}\frac{d\log g_*}{d\log x}+\frac{n_D(n_D-1)}{12}\frac{g_*(x_0)}{g_*(x)}\eta\left(\frac{x_0}{x}\right)^{n_D-4}\right]\frac{m\langle\sigma v \rangle}{x^2} \left(Y^2-Y^2_{\rm eq}\right). \label{eq:singleCase}
\end{align}
\end{widetext}
The terms in the square bracket are often neglected because in the radiation dominated case, it is approximately 1. However, if $\rho_D$ dominates resulting in a large value of $\eta$, then it is clear that this term becomes an important contribution.

\section{Analytic Solutions}
\label{sec:Analytic}

In this section, we take an analytic approach to solving equation \ref{eq:general} in terms of $\rho$ and $w$. We start by defining $\Delta=Y-Y_{\rm eq}$ and combine it with Eq. \ref{eq:generalSimple}:
\begin{align}
\frac{d\Delta}{dx}=-\frac{dY_{\rm eq}}{dx}-\frac{s}{\dot{x}}\langle\sigma v\rangle\Delta\left(\Delta+2 Y_{\rm eq}\right).
\end{align}
For small values of $x$, $\Delta$ and $\frac{d\Delta}{dx}$ are small as $Y$ tracks $Y_{\rm eq}$ closely. Using these approximations, as well as equations \ref{eq:sdot} and \ref{eq:neq}, solving for $\Delta$ gives:
\begin{align}
\Delta=\frac{1}{s\langle\sigma v\rangle\left(\frac{\Delta}{Y_{\rm eq}}+2\right)}\left(\dot{x}+\frac{3}{2}\frac{\dot{x}}{x}-3H\right).
\end{align}
The criterion for freezeout is given by $\Delta(x_f)=cY_{\rm eq}$ for some $c$. This $c$ will define the freezeout temperature, the point at which we change from the high temperature limit solution to the low temperature limit solution. We give more details about the choice of $c$ in section \ref{sec:numAna} where we choose $c$ to best fit the numerical results. Using the early time solution, we can obtain an implicit equation for $x_f$:
\begin{align}
cn_{\rm eq}(x_f)=\frac{1}{(c+2)\langle\sigma v\rangle}\left(\dot{x}|_{x=x_f}+\frac{3}{2}\frac{\dot{x}|_{x=x_f}}{x_f}-3H(x_f)\right), \label{eq:findXf}
\end{align}
where $H$ is given in Eq \ref{eq:FriedmanSingle}, $\dot{x}$ is given in Eq \ref{eq:xdot}, and $n_{\rm eq}$ is given in Eq \ref{eq:neq}. Without knowing the functional forms of $\rho$ and $w$, this is as far as we can go in general. Even if Eq. \ref{eq:findXf} is a complicated function, it can always be solved numerically for a value of $x_f$ much  quicker than solving \ref{eq:general}.

Once $x_f$ is found, we can look at $\ref{eq:generalSimple}$ in the large $x$ case where $Y\approx \Delta$ and $Y_{\rm eq}\approx 0$:
\begin{align}
\frac{d\Delta}{dx}=-\frac{s}{\dot{x}}\langle\sigma v\rangle \Delta^2.
\end{align}
Integrating from $x=x_f$ to $x\rightarrow\infty$ gives:
\begin{align}
Y(x\rightarrow \infty)=\left(\left(cY_{\rm eq}(x_f)\right)^{-1}+\int_{x_f}^{\infty} \frac{s}{\dot{x}}\langle\sigma v\rangle dx\right)^{-1}, \label{eq:yinfty1}
\end{align}
where $Y_{\rm eq}=n_{\rm eq}/s$, $s$ is given in Eq \ref{eq:s}, $\dot{x}$ is given in Eq \ref{eq:xdot}, and $n_{\rm eq}$ is given in Eq \ref{eq:neq}. It is convenient to neglect the contribution from $\left(cY_{\rm eq}(x_f)\right)^{-1}$ as doing so ensures that $Y(x\rightarrow \infty)$ is a strictly increasing function of $c$. Not ignoring this term can lead to numerical issues such as having two values for $c$ orders of magnitude apart which give the correct relic density, or there being no values of $c$ which give the correct relic density. This gives:
\begin{align}
Y(x\rightarrow \infty)=\left(\int_{x_f}^{\infty} \frac{s}{\dot{x}}\langle\sigma v\rangle dx\right)^{-1}, \label{eq:yinfty}
\end{align} 
Using Eq. \ref{eq:yinfty}, we can find the final relic density:
\begin{align}
\Omega h^2=\frac{m s_0 Y(x\rightarrow \infty)h^2}{\rho_C},
\end{align}
where 
\begin{align}
s_0=2970\ \mbox{cm}^{-3},
\end{align} 
is the entropy density today and 
\begin{align}
\rho_C =\frac{3 H_0^2}{8\pi G}\approx \left(1.054\times 10^{-5} \frac{\mbox{GeV}}{\mbox{cm}^3}\right)h^{2},
\end{align}
is the critical energy density today. Here, $H_0$ denotes the value of the Hubble parameter today and is usually given by $H_0=100h \mbox{ km}/\mbox{s}/\mbox{Mpc}\approx 70\mbox{ km}/\mbox{s}/\mbox{Mpc}$ \cite{Kolb:1990vq}.

\section{Numerical Analysis}
\label{sec:numAna}

\subsection{Solving the Boltzmann equation numerically}

In this section, we will investigate numerical solutions to equation \ref{eq:general}. To do so, we will assume that the Dark Matter is composed of scalars and that the thermally averaged cross section is approximately constant with
\begin{align}
\langle \sigma v \rangle = \sigma_0.
\end{align}
We will assume that the Dark Matter was in thermal equilibrium with the Standard Model well before it froze out. \footnote{If we assumed the Dark Matter were fermions, we would just change the number of degrees of freedom from 1 to 2 for Majoranna fermions or 4 for Dirac fermions. However, this does not qualitatively change the results. }

The parameters that we are interested in varying are the mass of the Dark Matter ($m$), the value of $\eta$, and the thermally averaged cross section $\sigma_0$, as well as investigating the cases of $n_D=6,8$. The $n_D=6$ scenario corresponds to some quintessence models with a kination phase while the $n_D=8$ corresponds to brane world cosmology or some late inflaton decay models \cite{Iminniyaz:2018das}. We are not interested in changing $x_0$ because any such change can be absorbed into a change in $\eta$. For this numerical analysis, we took $x_0=25$.

\begin{figure}[h]
\begin{center}
$
\includegraphics[scale=0.20,keepaspectratio=true]{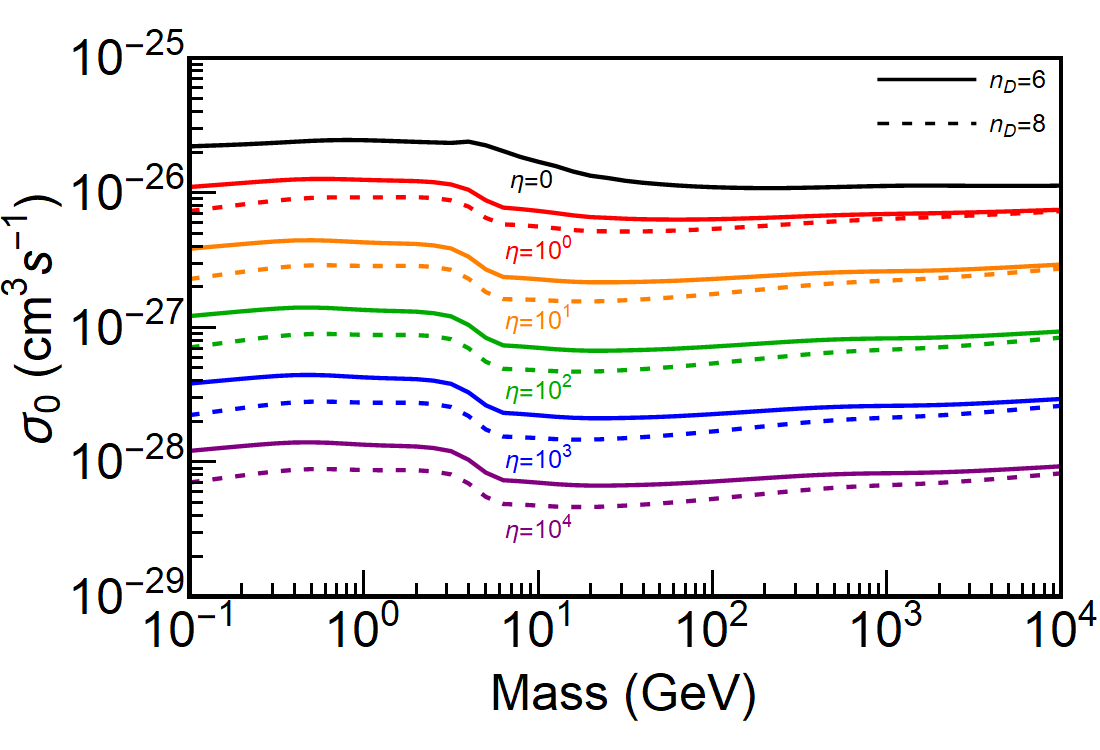} 
$ 
\end{center}

\vskip -0.1in
   \caption{ The required thermally averaged cross section to obtain the observed relic density $\Omega_{\rm CDM}h^2=0.120$ as a function of mass. The black, red, orange, green, blue, and purple (from the top to the bottom) curves represent $\eta=0, 1 ,10, 10^2, 10^3, 10^4$, respectively. The solid colored curves represent $n_D=6$ while the dashed curves represent $n_D=8$. These coincide for the black (top) curve.
}
\label{fig:sigVsM}
\end{figure}

In figure \ref{fig:sigVsM}, we show the required thermally averaged cross section to obtain the observed relic density $\Omega_{\rm CDM}h^2=0.120$ for various values of $\eta$. First, we should comment on the shape of the $\eta=0$ curve, namely why it starts by increasing, followed by a sharp decrease, followed by another increase. It should be noted that the typical value of $x$ for freeze-out is in the range $x=10$ to $x=30$, so the important temperatures to consider are on the order of $m/10$. Since $\Omega_{\rm CDM}\sim mn$ where $m$ is the Dark matter mass and $n$ is the Dark Matter number density, we see that the general trend for increasing the mass should be to decrease the number density $n$ by increasing the thermally averaged cross section. In the low mass and high mass regimes, this is what happens since $g_*(x)$ does not change significantly for the important values of $x$. However, around $m=50$ to $m=200$ GeV, the Dark Matter is freezing out around the same time as the QCD phase transition, resulting in a rapidly changing $g_*(x)$. The increased numbers of degrees of freedom as we increase the mass results in requiring a decrease in the thermally averaged cross section to obtain the correct relic density. The derivative term $\frac{1}{3}\frac{d\log g_*}{d\log x}$ is small at all temperatures and does not significantly influence the above trends. Although it is usually ignored, we include it in our analysis.

As expected from equation \ref{eq:general}, we see from figure \ref{fig:sigVsM} that increasing the value of $\eta$ results in requiring a smaller thermally averaged cross section. Even modest values of $\eta$ can have large effects on this value. We also see that the $n_D=8$ case always results in needing a smaller thermally averaged cross section.

\begin{figure}[h]
\begin{center}

\includegraphics[scale=0.20,keepaspectratio=true]{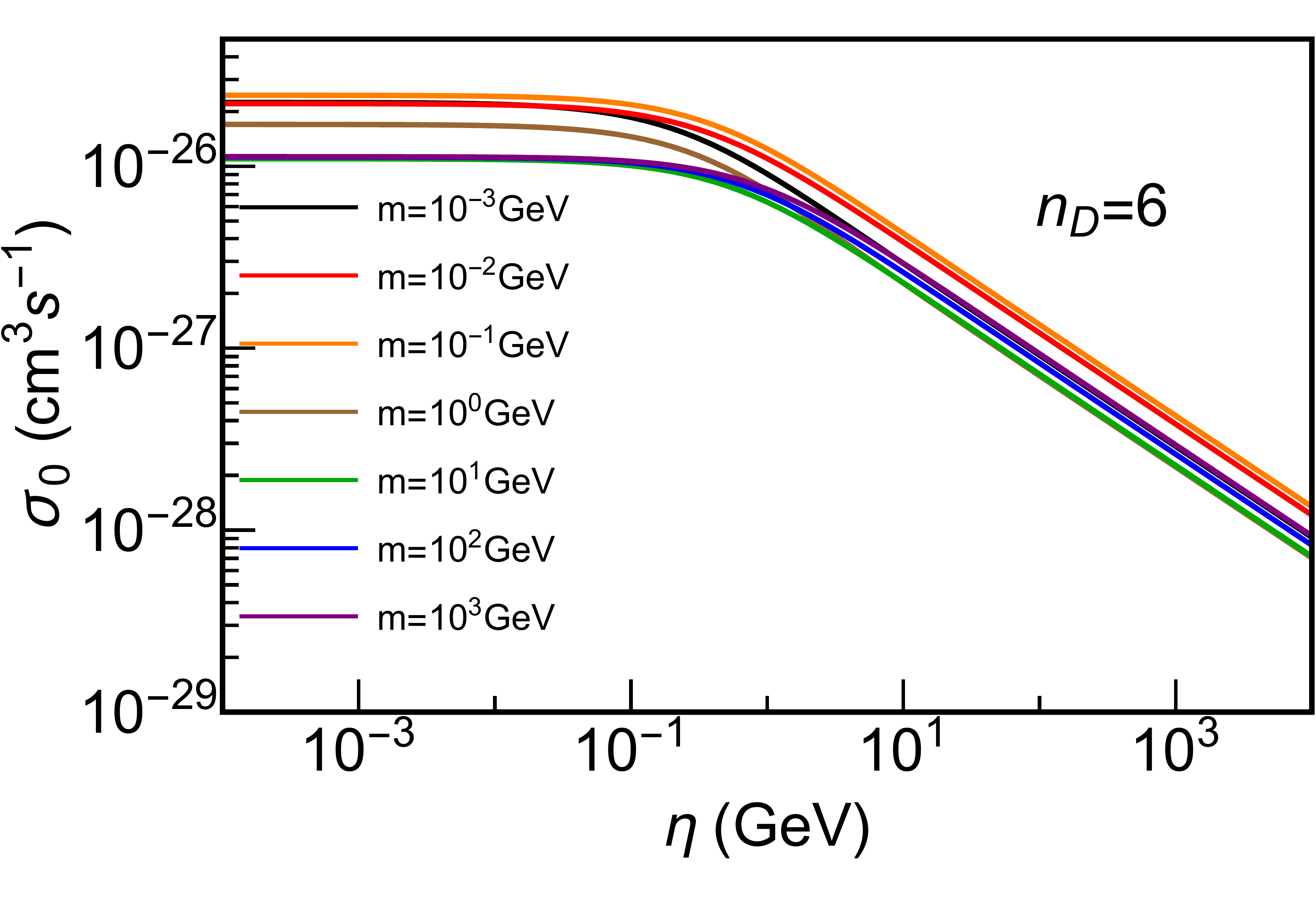} 
\includegraphics[scale=0.20,keepaspectratio=true]{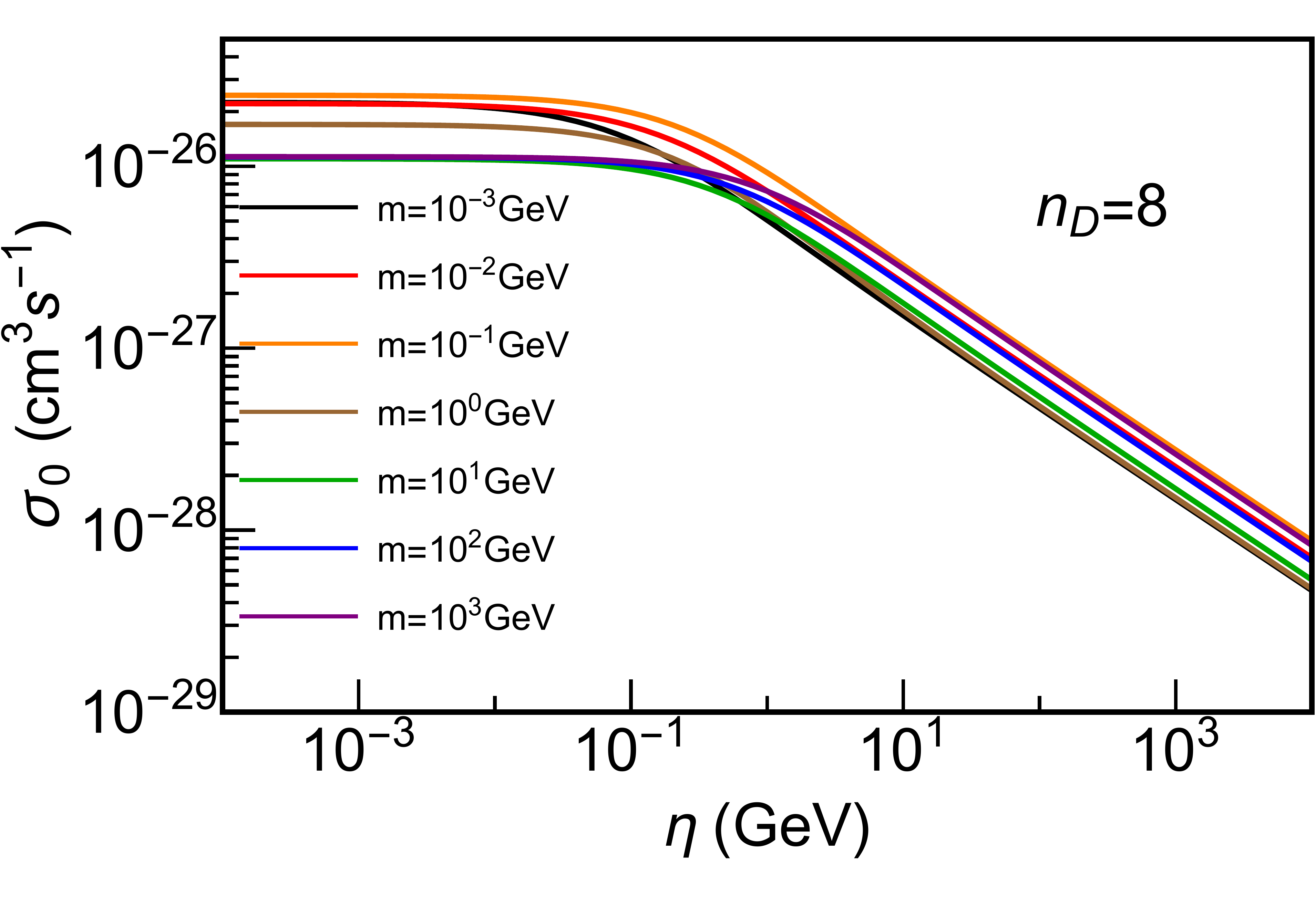}

\end{center}

\vskip -0.1in
   \caption{ The required thermally averaged cross section as a function of $\eta$ to obtain the correct relic density $\Omega_{\rm CDM}h^2=0.120$. The top figure is for $n_D=6$ while the bottom figure is for $n_D=8$. The various colors black, red, orange, brown, green, blue, and purple represent masses from $10^{-3}-10^3$ GeV by increments of an order of magnitude, respectively. 
}
\label{fig:sigVsEta}
\end{figure}

In figure \ref{fig:sigVsEta}, we show the required thermally averaged cross section as a function of $\eta$ to obtain the correct relic density $\Omega_{\rm CDM}h^2=0.120$ for various masses. One thing to notice is that the masses do not follow any nice pattern in terms of where they start off for low $\eta$. This is understood by imagining picking points from the $\eta=0$ curve in figure \ref{fig:sigVsM} at various masses and projecting them onto the $\sigma_0$ axis. Because of the irregular shape of the curve, the starting points for the masses will also be irregular.

\begin{figure}[h]
\begin{center}

\includegraphics[scale=0.20,keepaspectratio=true]{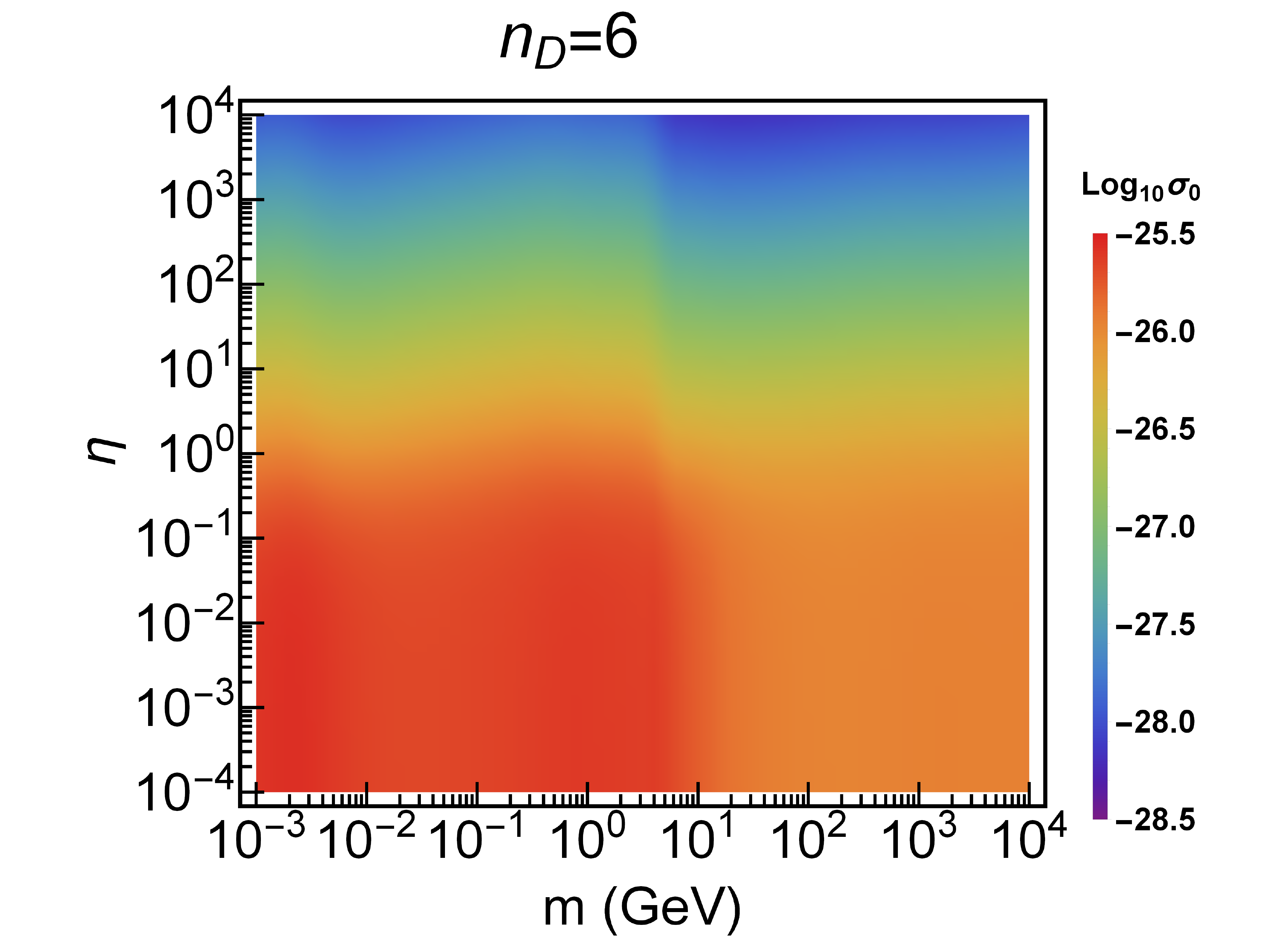} 
\includegraphics[scale=0.20,keepaspectratio=true]{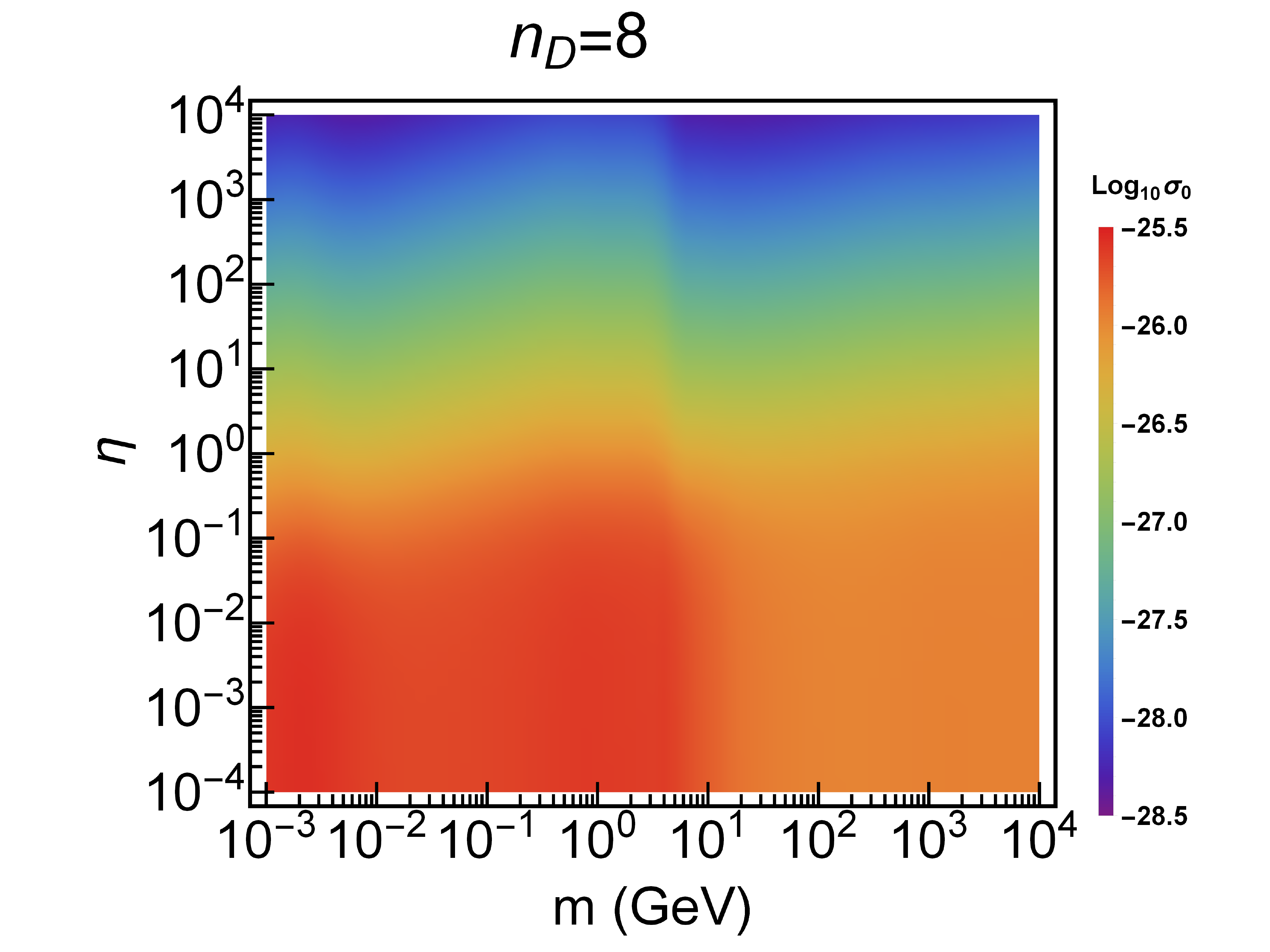}

\end{center}

\vskip -0.1in
   \caption{ The thermally averaged cross section required to obtain the correct relic density $\Omega_{\rm CDM}h^2=0.120$ as a function of mass and $\eta$. The top figure represents $n_D=6$ while the bottom figure represents $n_D=8$. The scale represents $\log_{10}\sigma_0$. 
}
\label{fig:mVsEta}
\end{figure}

\begin{figure}[h]
\begin{center}

\includegraphics[scale=0.20,keepaspectratio=true]{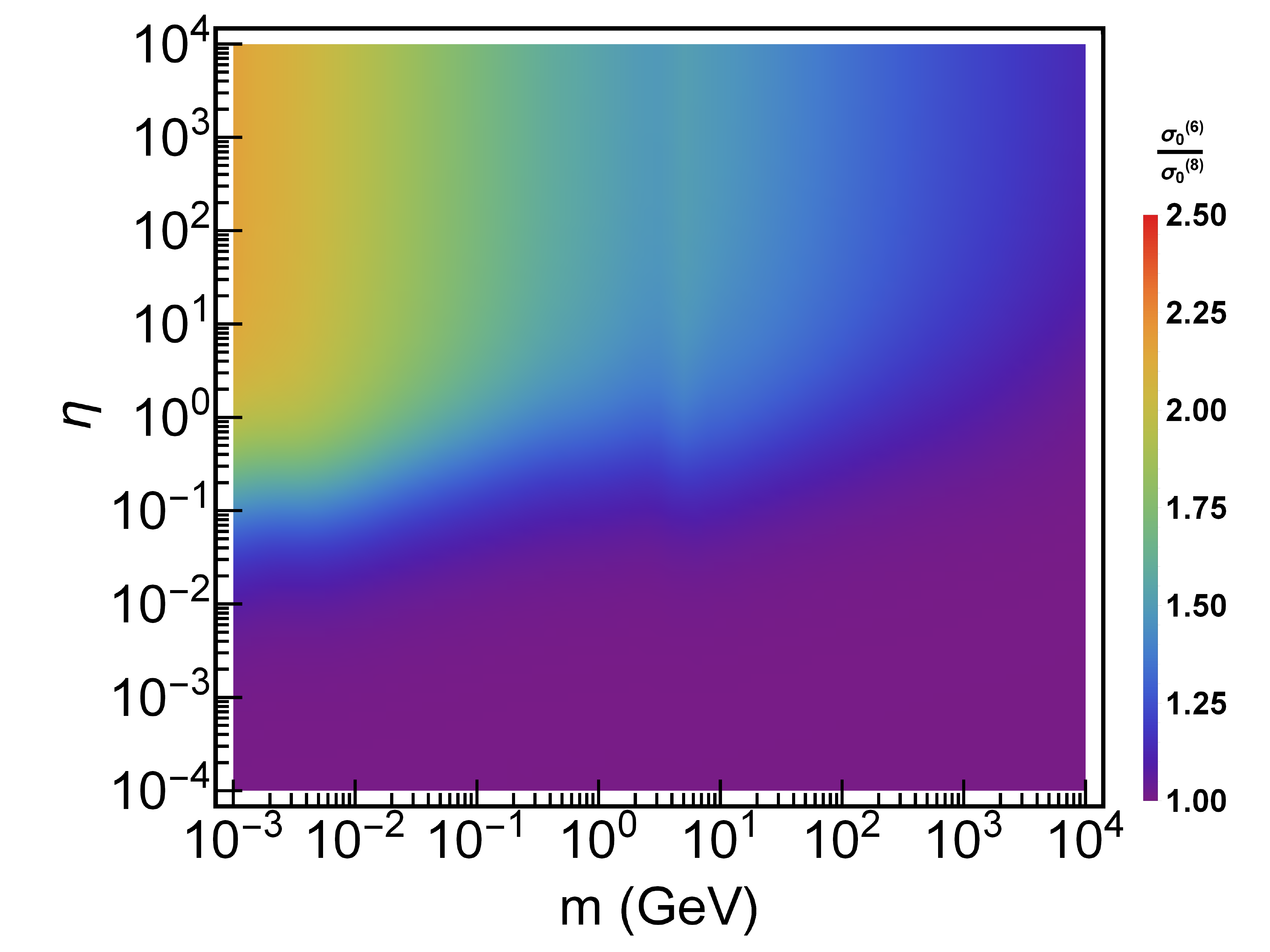} 

\end{center}

\vskip -0.1in
   \caption{ The ratio of the thermally averaged cross section of the $n_D=6$ and $n_D=8$ cases required to obtain the correct relic density $\Omega_{\rm CDM}h^2=0.120$ as a function of mass and $\eta$. $\sigma_0^{(6)}$ represents the cross section for $n_D=6$ while $\sigma_0^{(8)}$ represents the cross section for $n_D=8$. 
}
\label{fig:mVsEtaRat}
\end{figure}

One thing we do see is that at very small $\eta$, there is no effect as expected, but at around $\eta=0.1$ for $n_D=6$ and $\eta=0.01$ for $n_D=8$, the effects of the modified expansion start to be seen. Between $\eta=0.1$ and $\eta=10$ for both cases, we see that the required thermally averaged cross section starts to decreases according to a power law. From equation \ref{eq:singleCase}, we can deduce that this power law will have the form
\begin{align}
\sigma_0\propto\eta^{-1/2}.
\end{align}

Figure \ref{fig:sigVsM} and \ref{fig:sigVsEta} give us an idea of how the cross section needs to change to get the correct relic density. In figure \ref{fig:mVsEta}, we show a full map which includes both effects. The various colors represent the value of $\log_{10}\sigma_0$. Here, we see that the cross section gets much smaller as $\eta$ increases and does not differ much as the mass increases. At first glance, it may be difficult to see any obvious differences between the $n_D=6$ case on the top and the $n_D=8$ case on the bottom. In figure \ref{fig:mVsEtaRat}, we plot the ratio $\sigma_0^{(6)}/\sigma_0^{(8)}$ where the superscript indicates the value of $n_D$. This shows us that the differences are more important at low mass and only differs by a factor of order 1.

At this point it should be noted that if the cross section required to obtain the correct relic density becomes too small, then it will bring the thermal equilibrium assumption into question. Although we do not worry about that in our case, if one was to consider much larger values of $\eta$, one would need to keep this in mind. Furthermore, our analysis will only hold if there is no entropy injection at the end of the new cosmological era defined by the dominance of the new energy density. If instead of a rapidly cooling energy density with $n_D>4$ we choose, for example, a new matter dominated era with $n_D=3$, the new energy density would need to decay away which would create an entropy injection diluting the DM. Further details about these scenarios can be found in \cite{Bernal:2018ins,Bernal:2018kcw,Hamdan:2017psw,Hardy:2018bph}.

\subsection{Comparison with analytic approach}

When deriving the analytic approach in section \ref{sec:Analytic}, the value of $c$ was ambiguous. The standard lore suggests that if $\langle \sigma v\rangle \approx \sigma_0 x^{-n}$ for $x\gtrsim 3$, then one should choose $c$ such that $c(c+2)=n+1$ \cite{Kolb:1990vq}. The value of $n$ is related to the velocity dependence of the cross section and, in practice, accounts for how fast the Dark Matter freezes out after it is out of equilibrium. The larger the value of $n$, the faster the species freezes out. Although we are taking $n=0$ in our s-wave approximations, the effect of the extra energy density gives the equation a term which looks like $n=n_D/2-2$. Depending on the value of $\eta$, this term will change what we should choose for $c$ to a point that it is not clear what should be chosen.

\begin{figure}[h]
\begin{center}

\includegraphics[scale=0.20,keepaspectratio=true]{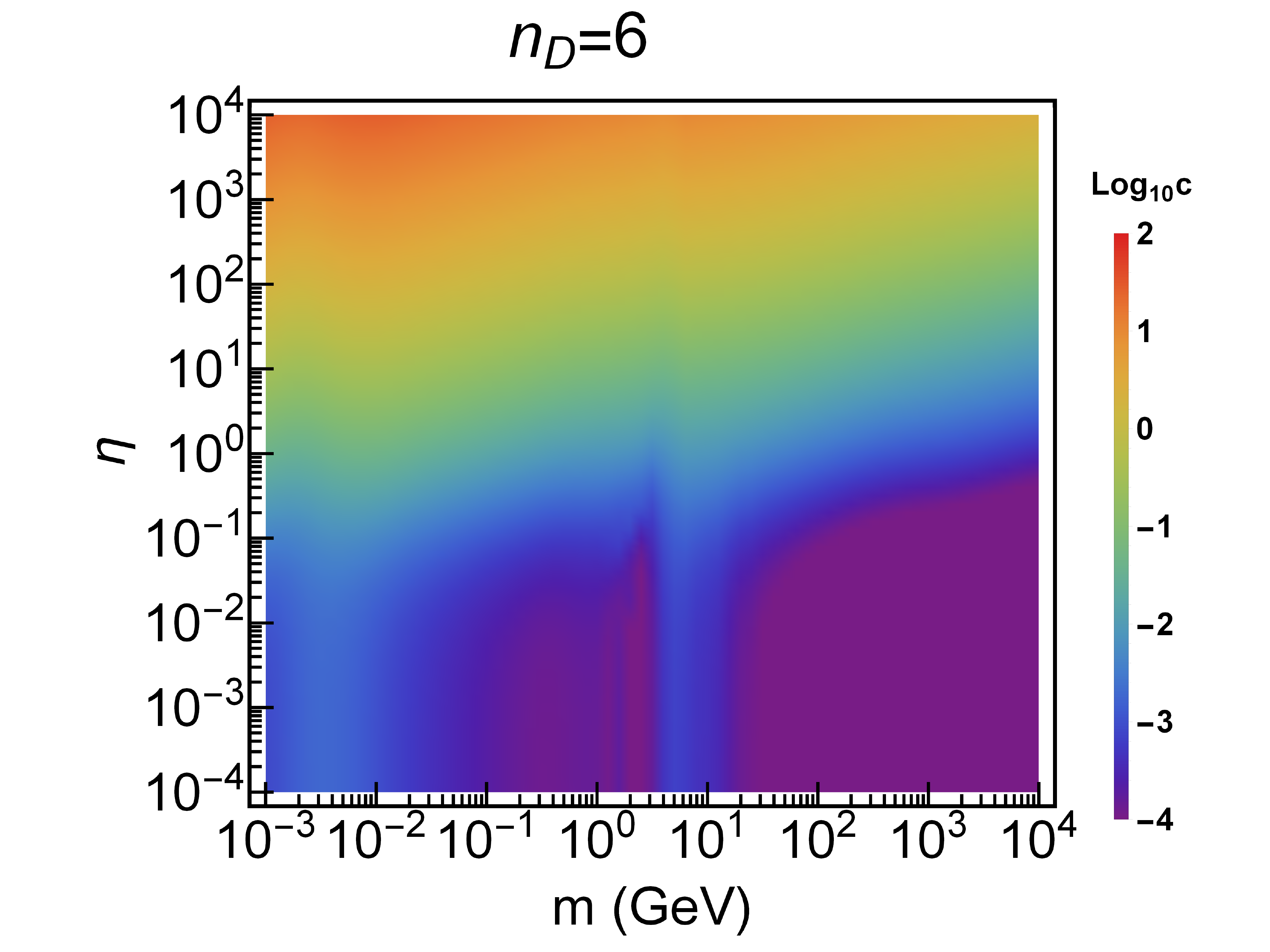} 
\includegraphics[scale=0.20,keepaspectratio=true]{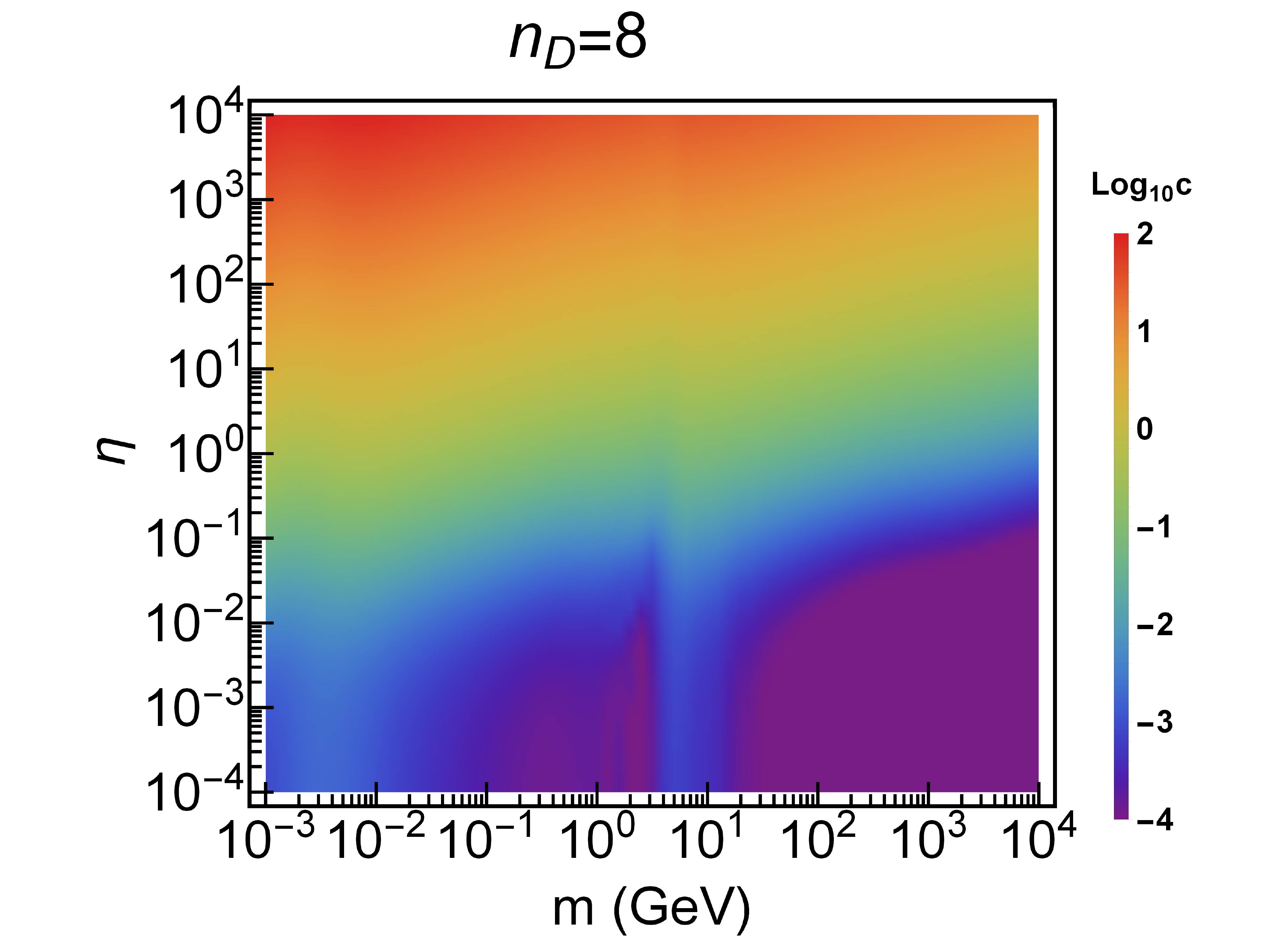}

\end{center}
\vskip -0.1in
   \caption{ The value of $c$ needed to obtained the correct relic density using the cross section values from figure \ref{fig:mVsEta}. The top plot has $n_D=6$ and the bottom plot has $n_D=8$. Plotted is the value of $\log_{10} c$. 
}
\label{fig:cVals}
\end{figure}

\begin{figure}[h]
\begin{center}

\includegraphics[scale=0.20,keepaspectratio=true]{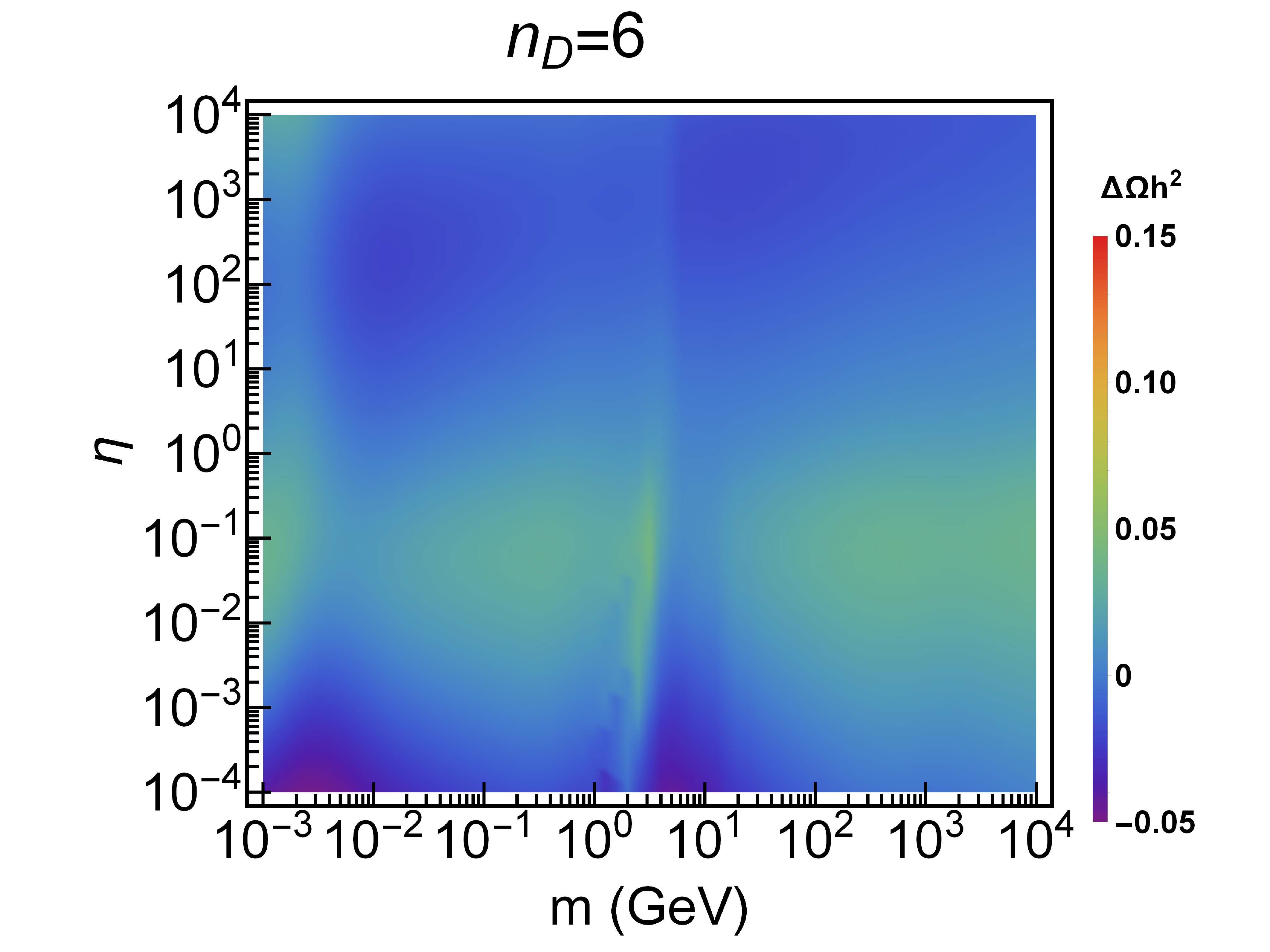} 
\includegraphics[scale=0.20,keepaspectratio=true]{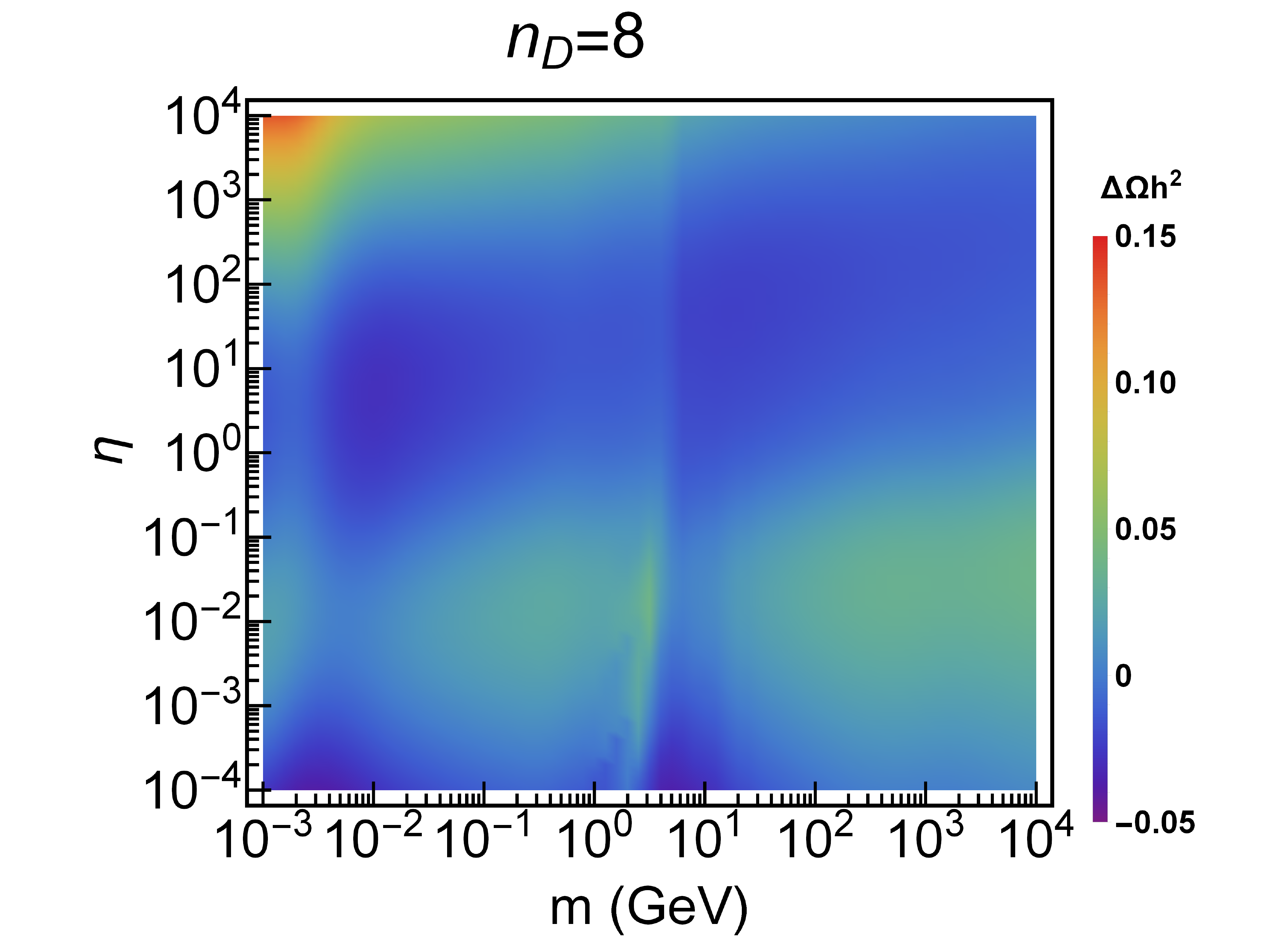}

\end{center}
\vskip -0.1in
   \caption{ The difference in the obtained relic density using the values of $c$ from Eq \ref{eq:cfunc} and the observed relic density. The top plot has $n_D=6$ and the bottom plot has $n_D=8$.
}
\label{fig:dOmega}
\end{figure}

We found that if we make fewer simplifications while using this approach, such as using Eq. \ref{eq:findXf} to find $x_f$ and Eq. \ref{eq:yinfty} to find $Y(x\rightarrow \infty)$ without neglecting terms, we need much different values of $c$ from the typical $c(c+2)=n+1$ to obtain the same relic density. See figure \ref{fig:cVals} for the values of $c$ needed to obtain the correct relic density using the cross section values obtained in figure \ref{fig:mVsEta}. We see that the value of $c$ changes drastically when we change $\eta$ but does not change much as the mass changes. Using this information, we can fit for $c$ to obtain the simple approximation:
\begin{align}
c\approx  \begin{cases} 
      0.0165m^{-0.219}\eta^{0.688} & \mbox{ for } n_D=6, \\
      0.0658m^{-0.233}\eta^{0.774} & \mbox{ for } n_D=8. 
   \end{cases}\label{eq:cfunc}
\end{align}
Using this approximation, we can calculate the relic density and see how it compares to the actual value. In figure \ref{fig:dOmega}, we plot
\begin{align}
\Delta \Omega h^2=\Omega_{\rm approx}h^2-0.120.
\end{align}

This approach works well for most of the parameter space explored but can be off by a factor of order one in some places. In particular at low $\eta$ and low mass, the analytic approximation tends to be too small while at high eta and low mass, the result $n_D=8$ tends to be too large. This does allow one to have the correct order of magnitude for the relic density which, depending on the purpose, can be good enough. In the range considered, the biggest deviations in the $n_D=6$ case were $\Omega h^2=0.0715$ on the low end and $\Omega h^2=0.161$ on the high end. For the $n_D=8$ case, the biggest deviations were $\Omega h^2=0.0812$ on the low end and $\Omega h^2=0.255$ on the high end.

One may be concerned about the large range of $c$ obtained and the validity of the approximations made in section \ref{sec:Analytic}. Recall that in the small $x$ regime, we assumed that $\Delta=Y-Y_{\rm eq}$ was small compared to $Y_{\rm eq}$ and for large $x$, we assume that $Y_{\rm eq}$ is much smaller than $\Delta$. Inevitably, when we go from one to the other, the assumption will break down and both values will be comparable. Where to choose this point becomes a question of what best fits the data.

\section{Summary and Conclusions}
\label{sec:conclusions}

In this work, we derived a general Boltzmann equation which can be used for any modified expansion history of the universe. This was done by characterizing the energy density as a general function and letting $w$ from the equation of state vary with temperature. We then looked at the specific example where we could write the energy density as a sum of components where all the values of $w$ from the equations of state for each component were constant. Finally, we looked at the case of a single extra energy density component parameterized by how fast it cools, $n_D$, and how abundant it was compared to radiation, $\eta$. Using these results, we found an analytic approach to approximating the relic density.

We then solved the modified Boltzmann equation for this last case to compare the effects to the standard picture. We found that for $n_D=6$ and $n_D=8$, the required cross section to obtain the measured relic does not changes significantly for $\eta\lesssim 10^{-1}$ and follows a power law for $\eta\gtrsim 10$ with some transition region between them. Importantly, a larger value of $\eta$ required a smaller cross section. This is quite simple to understand if we were to imagine what the expansion history of the universe would be if we ignored radiation. In this hypothetical situation, the scale factor would go as $a\propto t^{2/n_D}$, which gives a Hubble expansion rate of $H=2 t/n_D$, meaning that at the same temperature, the universe would be expanding slower than the radiation dominated case where $H=t/2$. Because of this, it takes more time for the rate of expansion to be comparable to the rate of the reaction keeping the Dark Matter in equilibrium, so freezeout would occur later. A smaller cross section for the equilibrium reaction, which normally results in the Dark Matter freezing out earlier, is required to balance this effect.

We also showed that in the $n_D=8$ case, the cross section needed is comparable to the $n_D=6$ case. The only significant difference being in the low mass, high $\eta$ case where they start to differ by a factor of order 1. 

This result has some important consequences for model building and Dark Matter detection. If one of these modified cosmological scenarios occurred in our Universe and Dark Matter went through freezeout, then it may well be the case that at weak scale masses, the cross sections are orders of magnitude smaller than current bounds. However, this opens the possibility that heavier Dark Matter particles underwent freezeout. The bound of the mass of Dark Matter undergoing freezeout is about $100$ TeV from perturbative unitarity \cite{Battaglieri:2017aum}, but this bound could be relaxed in these scenarios. This again poses a problem for Dark Matter detection as they are not designed to look for Dark Matter at these masses. All in all, these results suggests that Dark Matter detection may be a bigger challenge than previously expected if there is a significant changes to the expansion rate of the early Universe.

\begin{acknowledgments}
The author thanks Stephen Godfrey, Catarina Cosme, and Maíra Dutra for guidance and helpful conversation. The author also thanks Keith Dienes and Brooks Thomas for introducing him to the subject of Dark Matter and freezeout. This work was supported by the Natural Sciences and Engineering Research Council of Canada.  
\end{acknowledgments}


\end{document}